\documentclass{agujournal2019}

\usepackage{lineno,layouts,microtype}

\newcommand{\navidcomment}[1]{{\color{red} [#1]}}

\usepackage{sansmath} 
\sansmath

\newcommand{\mathbfit}[1]{\textbf{\textit{\textsf{#1}}}}
\usepackage{amsfonts, amssymb, amsmath}
\usepackage[labelfont=bf]{caption}

\usepackage{comment}

\journalname{Geophysical Research Letters}

\usepackage{scalerel, stackengine}
\setstackEOL{\#}
\stackMath
\def\hatgap{2pt}
\def\subdown{-2pt}
\newcommand\reallywidehat[2][]{ \renewcommand\stackalignment{l} \stackon[\hatgap]{#2}{ \stretchto{
    \scalerel*[\widthof{$#2$}]{\kern-.6pt\bigwedge\kern-.6pt}
    {\rule[-\textheight/2]{1ex}{\textheight}}}
    {0.5ex}_{\smash{ \belowbaseline[\subdown]{\scriptstyle#1} }}
}}

\newcommand{\bu}		{\mathbfit u}

\newcommand{\ee}		{\mathrm{e}}

\newcommand{\half}		{\tfrac{1}{2}}

\newcommand{\av}[1]	    {\left \langle {#1} \right \rangle}
\newcommand{\beq}		{\begin{equation}}
\newcommand{\eeq}		{\end{equation}}

\newcommand{\Pa}		{\mathrm{N}\,\mathrm{m}^{-2}}

\newcommand{\rhom} {\rho_{\mathrm{m}}}
\newcommand{\ws} {\mathrm{WS}}
\newcommand{\tfs} {\mathrm{TFS}}
\newcommand{\ifs} {\mathrm{IFS}}
\newcommand{\bd} {\mathrm{BD}}
\newcommand{\hb} {h_{\mathrm{bot}}}
\newcommand{\pb} {p_{\mathrm{bot}}}

\hypersetup{
	breaklinks,
	colorlinks=true,
	linkcolor=Blue,
  urlcolor=Purple,
	citecolor=Purple,
	pdfauthor={N. C. Constantinou and A. McC. Hogg}
 }

\begin{document}
\justify
\title{Eddy saturation of the Southern Ocean:\\ a baroclinic versus barotropic perspective}

\authors{
Navid C. Constantinou\affil{1}\quad and\quad Andrew McC. Hogg\affil{1}
}

\affiliation{1}{Research School of Earth Sciences and ARC Centre of Excellence for Climate Extremes,\\ Australian National University, Australia}

\correspondingauthor{Navid C. Constantinou}{navid.constantinou@anu.edu.au}


\begin{keypoints}
\item An isopycnal layered model, with a varying number of fluid layers, is used to assess relative importance of barotropic and baroclinic processes in the Southern Ocean.
\item Both baroclinic and barotropic flows exhibit regimes in which mean zonal transport is insensitive to wind stress.
\item Eddies actively shape the time-mean flow, irrespective of the instabilities from which they originate. 
\end{keypoints}

%
%

\begin{abstract}
``Eddy saturation'' is the regime in which the total time-mean volume transport of an oceanic current is relatively insensitive to the wind stress forcing and is often invoked as a dynamical description of Southern Ocean circulation. We revisit the problem of eddy saturation using a primitive-equations model in an idealized channel setup with bathymetry. We apply only mechanical wind stress forcing; there is no diapycnal mixing or surface buoyancy forcing. Our main aim is to assess the relative importance of two mechanisms for producing eddy saturated states: \emph{(i)}~the commonly invoked baroclinic mechanism that involves the competition of sloping isopycnals and restratification by production of baroclinic eddies, and \emph{(ii)}~the barotropic mechanism, that involves production of eddies through lateral shear instabilities or through the interaction of the barotropic current with bathymetric features. Our results suggest that the barotropic flow-component plays a crucial role in determining the total volume transport.

\noindent \textbf{Plain language summary} \\
\noindent Wind stress at the surface of the ocean is an important driver of ocean currents. However, what sets up the strength of the currents remains puzzling. The strongest ocean current flows around Antarctica: the Antarctic Circumpolar Current (ACC). It is believed that the ACC is close to a so-called ``eddy saturated'' state, a regime in which changes in the strength of the wind stress forcing do not alter the strength of the mean current. Instead, the swirling oceanic eddy motions that accompany the current are enhanced. Here, we investigate the physics and assess the relative importance of the two mechanisms proposed in the literature to explain this phenomenon: the most commonly invoked baroclinic mechanism and the recently proposed barotropic mechanism that crucially involves the interaction of the oceanic flow with bathymetric features. Our results suggest that the oftentimes ignored depth-averaged (barotropic) component of the ocean flow and its interaction with bathymetry play a dominant role in setting up the strength of the current in certain regimes.
\end{abstract}

%
%

\section{Introduction}

The Southern Ocean, and in particular the Antarctic Circumpolar Current (ACC) that connects all  ocean basins, influences the global ocean circulation and the climate \cite{Toggweiler-etal-2006, Talley-2013, Ferrari-etal-2014}. The circulation in the Southern Ocean is fueled by a combination of the strong westerly winds imparting momentum at the ocean's surface and by surface buoyancy forcing. In the last few decades the strength of the westerly winds over the Southern Ocean is increasing as a response to climate-change forcing and ozone depletion \cite{Marshall-2003,Swart-Fyfe-2012,Bracegirdle-etal-2013,Farneti-etal-2015}. Thus, significant effort has been expended to understand how the ACC will respond to this wind increase. However, several questions remain outstanding.

One idea that has emerged is that the ocean is in a so-called ``eddy saturated'' state, in which stronger winds do not increase the mean strength of the current. {\color{black}Instead, the additional work done by wind increases the energy of the mesoscale eddies.} This idea was first hypothesized by \citeA{Straub-1993}; since then a series of eddy-resolving ocean models, both idealized and more realistic, have verified that the ACC is close to the so-called ``eddy-saturated'' limit (e.g., \citeA{Hallberg-Gnanadesikan-2001,Tansley-Marshall-2001,Hallberg-Gnanadesikan-2006,Hogg-etal-2008,Farneti-etal-2010,Meredith-etal-2012,Dufour-etal-2012,Munday-etal-2013,Abernathey-Cessi-2014,Marshall-etal-2017}). {\color{black}The eddy-saturation hypothesis is not contradicted by observations \cite{Boning-etal-2008, Firing-etal-2011, Hogg-etal-2015}.}

In a seminal paper \citeA{Munk-Palmen-1951} argued that zonal momentum in the Southern Ocean is balanced primarily through `topographic form stress', rather than by bottom drag. Topographic form stress is an inviscid mechanism that couples the ocean to the solid Earth through correlations between the bathymetric slopes and the bottom pressure. Since the work by \citeA{Munk-Palmen-1951} it has been recognized that bathymetry and topographic form stress play a key role in setting the ACC transport.

Similar to topographic form stress, `interfacial form stress' describes the coupling between layers of different density and is responsible for the vertical transfer of momentum between those layers \cite{Johnson-Bryden-1989,Olbers-etal-2004,Ward-Hogg-2011}. In this sense, interfacial form stress requires variation of density in the vertical. The usual notion, which goes back to \citeA{Johnson-Bryden-1989}, is that at equilibrium interfacial form stress transfers eastward momentum downwards from the surface to the bottom of the ocean, where topographic form stress acts to transfer this momentum to the solid Earth. \citeA{Ward-Hogg-2011} demonstrated that when the wind stress at the ocean surface changes, rapid barotropic signals form (within several days); this induces topographic form stress that balances a large fraction of the imparted momentum from the surface directly to the bottom.\footnote{Such a fast barotropic response to changes in the wind can be seen in the Southern Ocean State Estimate \cite{Masich-etal-2015} and it is expected; the wind stress projects directly onto the barotropic mode.} {\color{black}Therefore, a corollary stemming from the work by \citeA{Ward-Hogg-2011} is that the} equilibrium of the sort envisaged by \citeA{Johnson-Bryden-1989} is established when downwards transfer of eastward momentum from the surface balances upwards transfer of westward momentum from the topography.

The most common explanation for how eddy-saturated states are established relies on the generation of eddies through baroclinic instability (see, e.g., \citeA{Straub-1993, Nadeau-Straub-2012, Marshall-etal-2017}). The role of bathymetry in setting up the ACC transport is acknowledged in this explanation, although the role of bathymetry in eddy generation is unclear. However, isolated bathymetric features can have a large effect on the transient (time-dependent) eddy field through localized baroclinic instability and also {\color{black}by producing} an associated almost-barotropic standing (time-mean) eddy field pattern \cite{Abernathey-Cessi-2014}.

The baroclinic view of eddy saturation is consistent with the notion that ACC transport is dominated by the `thermal-wind transport' -- the transport that is diagnosed from the density field alone{\color{black}, i.e., by integrating the thermal-wind balance twice} assuming zero velocity at the bottom of the ocean. Thus, thermal-wind transport can be easily inferred from hydrography measurements of the density. Furthermore, the thermal-wind transport is related to baroclinic instability, since the vertical shear  fuels production of eddies through baroclinic instability. Therefore, considerable effort has been put in explaining why the `thermal-wind' transport of the ACC is saturated (see, e.g., \citeA{Marshall-etal-2017}).

However, the notion of zero mean flow at the bottom of the ocean (as the thermal-wind transport assumes) has been, recently, challenged. \citeA{Donohue-etal-2016} made detailed measurements at the ocean floor of Drake Passage (cDrake experiment) and found time-mean\footnote{Time-mean calculated as the average over the four years that the cDrake experiment was in operation.} bottom flows as strong as about 0.1$\,\text{m}\,\text{s}^{-1}$. Thus, the bottom flow can substantially contribute to the total transport. Characteristically, the results by \citeA{Donohue-etal-2016} resulted in the `nominal' value of the Drake Passage transport increasing from around 130$\,\text{Sv}$ to about 170$\,\text{Sv}$ ($\text{Sv}=10^6\,\text{m}^3\,\text{s}^{-1}$).  Similar findings regarding the importance of bottom flows in the Southern Ocean were also made from satellite altimetry and output from the Southern Ocean State Estimate \cite{Rintoul-etal-2014, PenaMolino-etal-2014, Masich-etal-2015}. These findings argue that, although the focus has been for long centered around the thermal-wind transport, the barotropic component of the flow may play an important role in setting up the ACC transport. 

In addition to the evidence for strong bottom flows, some recent work has emphasized the importance of the bathymetry in setting up standing meanders in the course of the ACC {\color{black}and in producing strong interaction between barotropic and baroclinic fluid motions} \cite{Youngs-etal-2017, Barthel-etal-2017}. Furthermore, these meanders play a crucial role in balancing the momentum through topographic form stress and thus determining the ACC transport \cite{Thompson-NaveiraGarabato-2014,Katsumata-2017}. These papers further argue for the importance of the barotropic mode in setting up the strength of the transport. Recent work has highlighted that eddy-saturated states can be obtained even without any baroclinicity, i.e., in a barotropic ocean of constant density \cite{Constantinou-Young-2017, Constantinou-2018}. For the establishment of this `barotropic eddy saturation' the bathymetry plays a crucial role. Without any baroclinicity the ocean must rely on instabilities due to lateral shear of the flow or due to the interaction of the flow with bathymetric features \cite{Hart-1979, Charney-Flierl-1980}. 

In this work, we examine the relative importance of the barotropic and baroclinic processes in establishing an eddy-saturated transport. We use a primitive-equations model in an idealized zonally re-entrant channel that is forced solely by wind stress. {\color{black}There is no diapycnal mixing or surface buoyancy forcing.} With this setup we depart from the quasi-geostrophic approach taken by \citeA{Constantinou-2018}, while our model remains computationally tractable  to enable us to span a wide range in parameter space. An important advantage of using an isopycnal layered model is that we can vary the stratification in a self-consistent manner that allows us to identify the contributions of baroclinicity in setting up the mean transport.

\section{Setup}

\begin{figure}[t]
\centering
\noindent\includegraphics[width=0.85\textwidth]{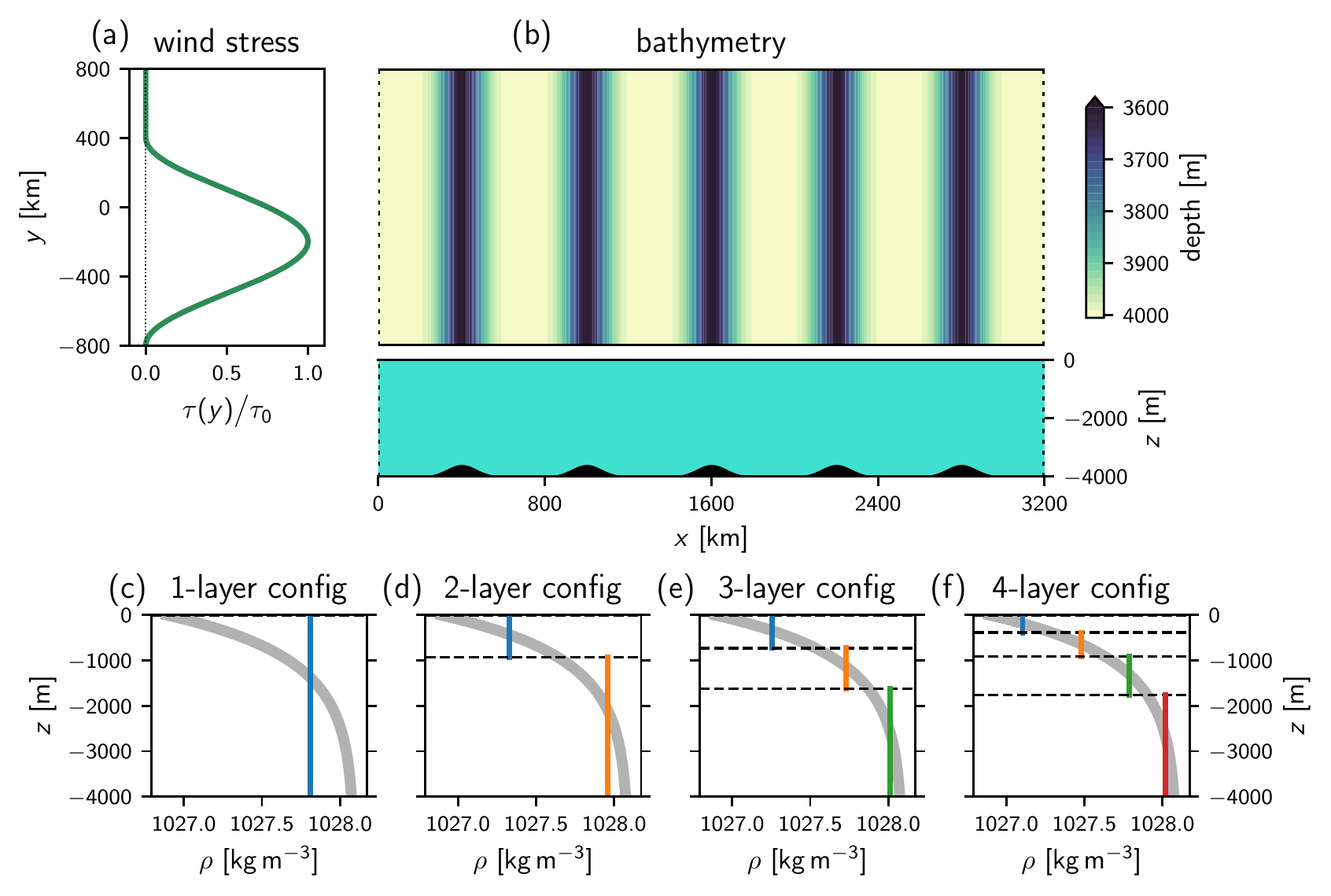}
\vspace*{-1em}\caption{(a) The meridional structure of the imposed wind stress. (b) The bathymetry; top~($x$--$y$) and side ($x$--$z$) view. (c)-(f) The layered discretizations of the reference density $\rho(z)$ (shown in solid gray). Dashed horizontal lines depict the rest-heights of the various fluid layers while vertical solid lines denote the mean density of the corresponding fluid layer.}\label{fig:domain}\vspace*{-4em}
\end{figure}

We use the Modular Ocean Model version~6 (MOM6)~\cite{Adcroft-etal-2019-mom6} to solve the primitive equations in isopycnal coordinates under the Boussinesq approximation and with a free surface. The model is set up in a zonally re-entrant channel on a beta-plane. The channel is $3200\,\mathrm{km}$ long in the zonal direction, $1600\,\mathrm{km}$ wide in the meridional direction, $4\,\mathrm{km}$ deep, and contains several Gaussian ridges with maximum height of $400\,\mathrm{m}$ and full-width-at-half-maximum of $165\,\mathrm{km}$; see figure~\ref{fig:domain}(b) for details. The horizontal grid cells are squares with $4\,\mathrm{km}$ grid spacing. The Coriolis parameter is $f=f_0+\beta y$ with $f_0=-10^{-4}\,\mathrm{s}^{-1}$ and $\beta=1.5\times 10^{-11}\,\mathrm{m}^{-1}\,\mathrm{s}^{-1}$; these values are typical of the Southern Ocean. We include frictional quadratic bottom drag {\color{black}with coefficient 0.003}, biharmonic lateral viscosity {\color{black}with coefficient $1.5\times10^9\,\mathrm{m}^4\,\mathrm{s}^{-1}$}, and free-slip {\color{black}insulating} sidewalls. We force the model with an imposed steady eastward wind stress that is zonally symmetric and has a meridional structure 
\begin{linenomath*}
\begin{equation}
    \tau(y) = \tau_0 \sin^2 \left(\frac{\pi y}{3L_y/4}\right),\quad\text{for } 0\le y\le \frac{3}{4}L_y, \label{eq:windstress}
\end{equation} 
\end{linenomath*}
shown in figure~\ref{fig:domain}(a). 

A nominal value for wind stress magnitude over the Southern Ocean is about 0.15$\,\Pa$ \cite{Risien-Chelton-2008}. Given the idealization of the wind stress~\eqref{eq:windstress} (steady forcing; no time-variability) and also the difference in the latitudinal extent of the forcing, we span four orders of magnitude in our experiments: $10^{-3}\,\Pa\lessapprox \tau_0\lessapprox 10\,\Pa$. This approach ensures that we cover realistic parameter regimes while also pushing the system to its limits.

The density profile is assumed to follow an exponential form,
\begin{linenomath*}
\begin{equation}
\rho(z) = \rho_0+ \Delta\rho \, (1-\ee^{z/d}),\quad -H\le z\le 0,
\end{equation} 
\end{linenomath*} 
with $\rho_0=1026.89\,\mathrm{kg}\,\mathrm{m}^{-3}$, $\Delta\rho=1.20\,\mathrm{kg}\,\mathrm{m}^{-3}$, and $d=950\,\mathrm{m}$.  Using a least-squares-fit 
we discretize this continuous density profile into $n$ fluid layers, where $n=1,2,3,4$ (figures~\ref{fig:domain}\mbox{(c)-(f)}). 
The discretization is designed to minimize variation in the first Rossby radius of deformation  with the number of layers.
In each case the first Rossby radius is  $20\pm 1\,\mathrm{km}$; the continuous density profile~$\rho(z)$ implies a deformation radius of $19.3\,\mathrm{km}$ \cite{LaCasce-2012}. These values of Rossby radius lie within the range found across the Southern Ocean \cite{Chelton-etal-1998}. 

This self-consistent manner of increasing fluid layers to approach the continuous density structure, $\rho(z)$, allows us to isolate the barotropic dynamics {\color{black} when $n=1$ from the combined effect of barotropic and baroclinic dynamics when $n\ge 2$}. We then  investigate the effects of these dynamics on the structure of the flow and on the response of the mean current to wind stress forcing. In the following sections, we refer to the  density discretization as `$n$-layer configuration'. On the other hand, flow in the middle layer of the 3-layer configuration is referred to as `the layer 2 flow'.

\section{Results}

All simulations presented are well-equilibrated; {\color{black}the time series of kinetic energy in each layer and potential energy for each interface reached statistical stationarity}. The equilibration times differ from case to case. For example, barotropic single-layer runs reach statistical equilibrium about 5 years after starting from rest, while weakly forced ($\tau_0<0.01\,\Pa$) baroclinic runs need up to 500 years to spin up. After the flow equilibrates, we average (for at least 30 years) to obtain the time-mean fields.

\subsection{Total time-mean transport}\label{sec:transport}

\begin{figure}
\centering
\noindent\includegraphics{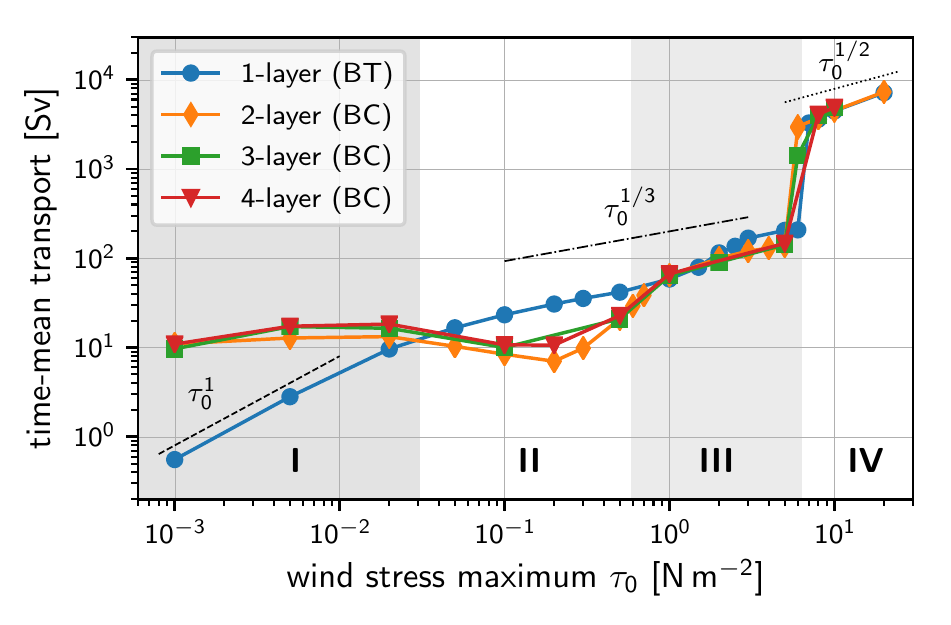}
\vspace*{-1em}\caption{The time-mean transport as a function of the wind stress forcing for the various barotropic (BT) and baroclinic (BC) configurations. Shadings mark the flow regimes discussed in section~\ref{sec:transport}.  ($\text{Sv} = 10^6\,\text{m}^3\,\text{s}^{-1}$)}\label{fig:transport}\vspace*{-3em}
\end{figure}

Figure~\ref{fig:transport} shows the equilibrated time-mean zonal transport of the channel as a function of the wind stress maximum ($\tau_0$) for each configuration. The total transport is computed as the time-mean of $\sum_k\int_0^{L_y} u_k h_k\,\mathrm{d}y$ evaluated at a fixed $x$, where $u$ is the zonal flow, $h$ is the layer depth, and subscript $k$ denotes each fluid layer with $k=1$ corresponding to the top layer.\footnote{For a detailed exposition of the primitive equations in isopycnal coordinates the reader is referred to the paper by \citeA{Ward-Hogg-2011}.}

The results from the barotropic runs (single-layer configuration) are qualitatively similar to those of \citeA{Constantinou-Young-2017} and \citeA{Constantinou-2018}. For very weak wind stress forcing ($\tau\le0.02\,\Pa$) the transport increases linearly with wind stress. For intermediate values of wind stress forcing ($0.05\,\Pa\le\tau_0\le 6.0\,\Pa$) the flow develops transient eddies and the transport becomes less sensitive to wind stress forcing. Here, during a 120-fold increase in wind stress the time-mean transport only increases 12-fold (from $17$ to $211\,\mathrm{Sv}$). Lastly, for higher values of the wind stress forcing ($\tau\ge 7.0\,\Pa)$ the flow abruptly transitions to a new regime with higher transport, in which the transport scales as~$\tau_0^{1/2}$ (shown in figure~\ref{fig:transport}).

The baroclinic configurations ($n\ge2$) show similar behavior among themselves and, in part, different behavior compared with the single-layer barotropic runs. For wind stress forcing values $\tau_0\le 0.5\,\Pa$ the time-mean transport remains practically insensitive to wind stress forcing; we even see a decrease of the transport with wind stress forcing (discussed further in section~\ref{sec:resultdetails1}). For $\tau_0 \ge 1.0\,\Pa$, the transport from the baroclinic configurations coincide with those from the single-layer barotropic runs, and follow the same transition to the bottom-drag-dominated, so-called, `upper-branch' regime.

In summary, figure~\ref{fig:transport} suggests that there exist four different regimes (marked I through~IV):
\begin{enumerate}
  \item[\textbf{I}.]  The baroclinic cases show {\color{black}weak sensitivity} on the total transport with wind stress; for barotropic runs the transport is lower  than the corresponding baroclinic cases and grows linearly with wind stress.
  \item[\textbf{II}.] The transport for barotropic cases exceeds that of baroclinic cases. The transport for barotropic runs grows at a rate much less than linear with wind stress; the transport for baroclinic runs decreases slightly.
  \item[\textbf{III}.] The baroclinic and barotropic cases show almost identical transports; these transports continue to grow at a rate much less than linear with wind stress.
  \item[\textbf{IV}.] The total transport both for barotropic and baroclinic cases undergoes an abrupt transition to much higher values (the `upper branch').
\end{enumerate}

Note that figure~\ref{fig:transport} only presents time-mean values. Although transport values for barotropic and baroclinic configurations are similar in certain flow regimes, the baroclinic flows are much more time-varying; more details on this in section~\ref{sec:resultdetails2}.

Since all baroclinic runs show qualitatively similar behavior we will, from hereafter, concentrate only on the purely barotropic 1-layer configuration and the baroclinic 2-layer configuration.

\subsection{Flow structure comparison}

\begin{figure}
\centering
\noindent\includegraphics{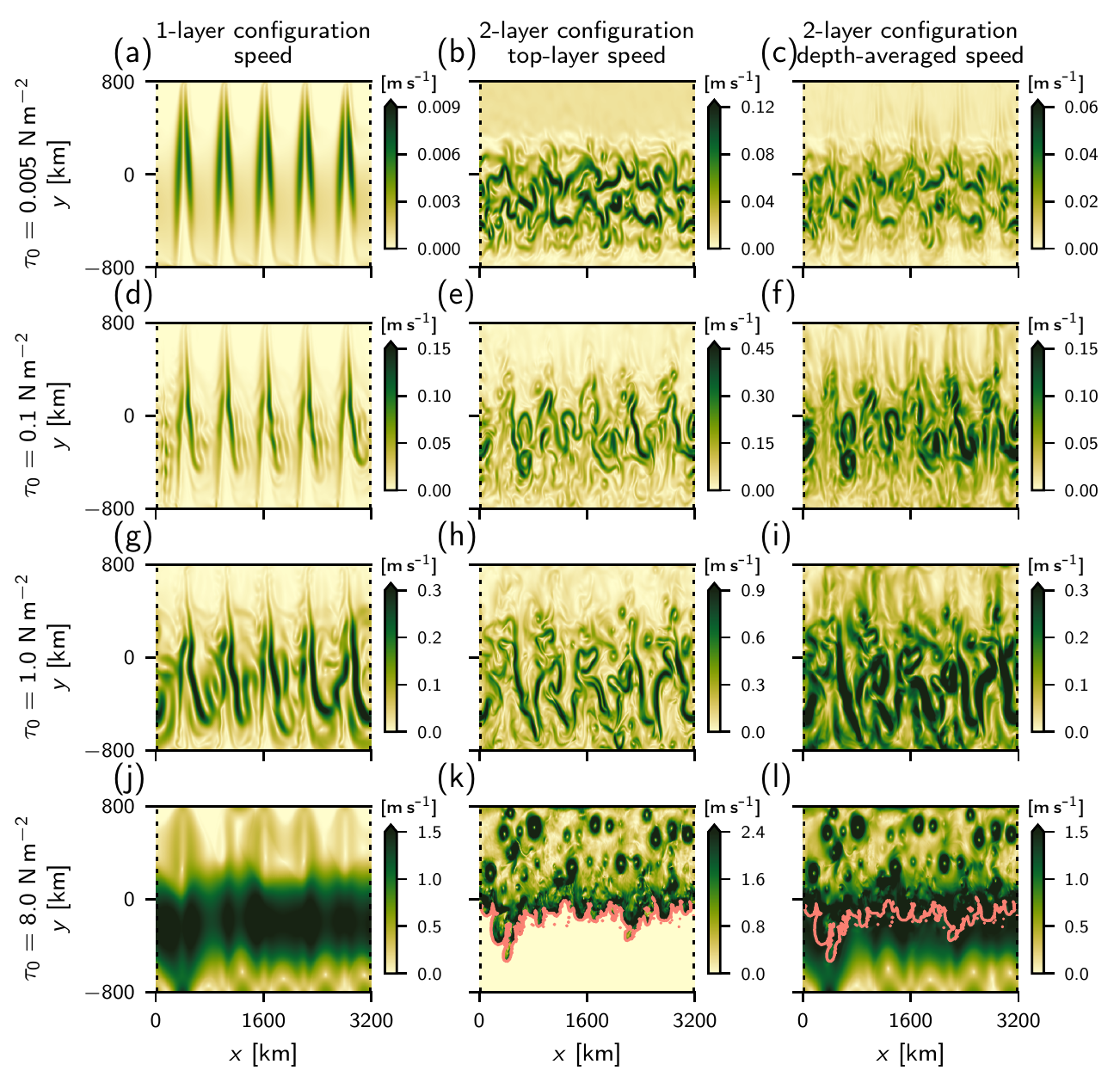}
\caption{Snapshots of the flow speed for 1-layer and 2-layer configurations and various wind stress forcing values. Left and middle columns show snapshots of the top-layer flow speed; right column shows the depth-averaged flow speed for the 2-layer configuration. In all baroclinic configurations ($n\ge 2$) the top-fluid layer outcrops when the flow is on the upper branch. The pink contour in panels~(k) and~(l) show the contour $h_1=0$, i.e., where the layer outcrops. Despite the dramatic difference in flow structure, the time-mean transport is the same {\color{black}between the 1-layer and 2-layer configurations} for cases with $\tau_0=1.0\,\Pa$ and $\tau_0=8.0\,\Pa$ {\color{black}respectively}.}\label{fig:snapshots}
\end{figure}

It is instructive to compare snapshots of statistically equilibrated flow for selected cases. Figure~\ref{fig:snapshots} shows flow speed for differing values of wind stress, one for each of the four different regimes described above. The left column of figure~\ref{fig:snapshots} shows snapshots from barotropic 1-layer configuration, while the middle and right columns show snapshots from baroclinic 2-layer configurations. Middle column shows top-layer speed, $|\bu_1|$, and right column shows the depth-averaged flow speed, $\sum_k h_k|\bu_k| \big/ h$, where $h=\sum_k h_k$ is the total depth. Note how different the flow structure is between the barotropic and baroclinic configurations, even for wind stress forcing $\tau_0\ge 1.0\,\Pa$ for which the total transport is identical amongst all layered configurations.

For weakly forced cases (the flow regime~I; figures~\ref{fig:snapshots}(a)-(c)) the barotropic flow is steady with no eddies and follows the geostrophic contours, $f/h$.  On the other hand, the baroclinic flow of figure~\ref{fig:snapshots}~(b,c) shows multiple jets and eddies that resemble {\color{black}homogeneous} baroclinic turbulence; the top-layer flow has a weak imprint of the bathymetry. In the flow regime~II (panels~(d)-(f)) the barotropic flow starts developing transient eddies but is steered by the bathymetry, while the baroclinic flow shows an imprint of the bathymetry itself. For wind stress values within the flow regime~III (panels~(g)-(i)) both barotropic and baroclinic flows show a strong imprint of the bathymetry while the baroclinic flow continues to be much more eddying. Finally, at the upper-branch flow regime~IV (panels~(j)-(l)) the barotropic runs develop a strong jet spanning all latitudes where wind stress is non-zero. The top-layer baroclinic flow is very different in this case: it outcrops at the surface (see panel~(k)) and is dominated by inertial eddies. These snapshots highlight the great differences in flow regimes that occur, despite the relative insensitivity of the time-mean zonal transport {\color{black}among the various layered configurations in flow regimes~III and~IV. A plausible rationalization for why higher wind stress values lead to similar time-mean transport for the various layered configurations is that more eddies imply more vertical momentum transfer through interfacial form stress leading to barotropization of the flow.}

The following subsections describe the different processes that are involved in determining the flow structure and the time-mean total transport (figures~\ref{fig:transport} and~\ref{fig:snapshots}).

\subsection{Transport per fluid layer and momentum balance}\label{sec:resultdetails1}

\begin{figure}
\centering
\noindent\includegraphics[width=\textwidth]{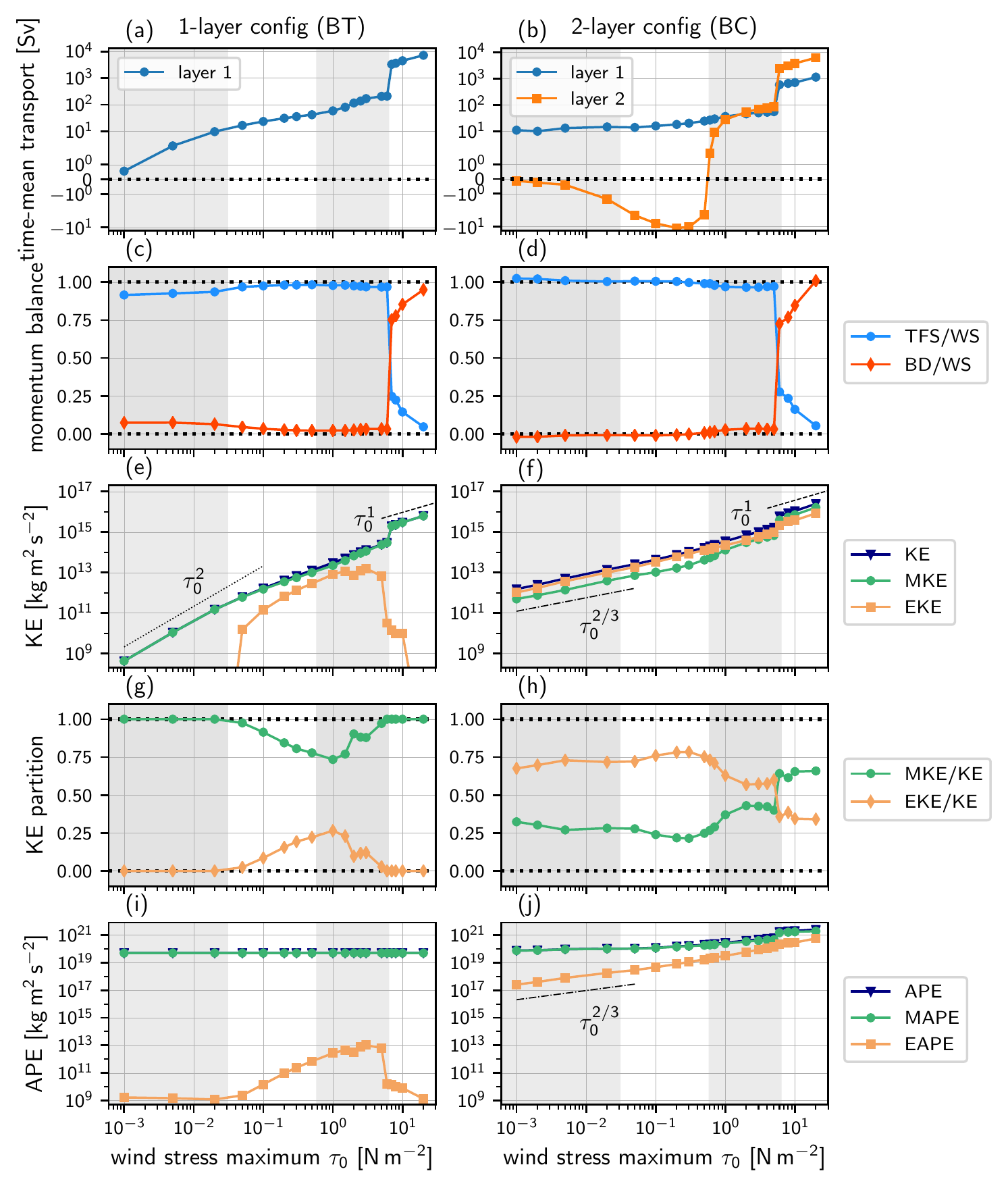}
\caption{(a)-(b)~The contribution to the time-mean transport from each fluid layer for the 1-layer and 2-layer configurations shown in figure~\ref{fig:transport}. (c)-(d)~The ratio of the integrated zonal momentum balanced by topographic form stress over the momentum imparted by wind stress, $\tfs/\ws$, and, similarly, the momentum balances by bottom drag, $\bd/\ws$. Note that in panel~(d) the bottom drag balance is actually slightly negative for $\tau_0\le 0.5\,\Pa$ as a result of the westward bottom-layer transport. (e)-(f)~The total kinetic energy (KE), the mean kinetic energy (MKE) and the eddy kinetic energy (EKE). (g)-(h)~The ratios  MKE$/$KE and EKE$/$KE. {\color{black}(i)-(j)~The total available potential energy (APE), the mean available potential energy (MAPE) and the eddy available potential energy (EAPE). } In all panels background shading separates the various flow regimes~I--IV described in figure~\ref{fig:transport}.}\label{fig:transportdecomp}\vspace*{-3em}
\end{figure}

First, let us see how the total time-mean transport for the baroclinic 2-layer runs  in figure~\ref{fig:transport} is split across fluid layers. Panels (a)-(b) of figure~\ref{fig:transportdecomp} show the decomposition of the time-mean transport into each fluid layer. What we immediately notice in panel~(b) is that for $\tau_0\le0.5\,\Pa$ the bottom layer has a westward (negative) time-mean transport. (We find such bottom-layer westward flows in all baroclinic runs for $n=2, 3, 4$ fluid layers; not shown.)

The depth integrated zonal momentum balance implies that the wind stress (WS) at the surface must be balanced by a combination of topographic form stress (TFS) and bottom drag (BD) in the bottom layer:
\begin{linenomath*}
\begin{equation}
\underbrace{\big\langle\,\tau\,\big\rangle}_{\ws} = \underbrace{\big\langle\, \overline{\pb\,\partial_x{\hb}}\,\big\rangle}_{\tfs} + \underbrace{\big\langle\, \rhom c_D\overline{|\bu_n|u_n}\,\big\rangle }_{\bd}, \label{eq:totalmombal}
\end{equation}
\end{linenomath*}
where angle brackets denote layer-average and overbar denotes time-average, $\hb$ is the bathymetry, and $\pb$ is the bottom pressure.

Figures~\ref{fig:transportdecomp}(c)-(d) show how the topographic form stress and bottom drag on the right-hand-side of~\eqref{eq:totalmombal} contribute to balance the wind stress on the left-hand-side. What stands out is that for most of the wind stress values, the main momentum balance is between wind stress and topographic form stress. Only in the upper-branch flow regime~IV does this momentum balance change. The flow in the upper branch barely feels the bathymetry, the topographic form stress becomes negligible, and bottom drag balances the wind stress resulting in very large time-mean transports. If the wind stress is to be balanced solely by the bottom drag then $\tau_0 \propto |\bu_n|u_n$ implying that transport should scale with $\tau_0^{1/2}$ as seen in figure~\ref{fig:transport}. {\color{black}Note, that both the `upper branch' flow regime~IV and the far left part of flow regime~I are not expected to be realized in the Southern Ocean, as they occur for unrealistically strong or weak values of the wind stress respectively.} The transition to the upper branch is precisely what was found previously in quasi-geostrophic single-layer simulations \cite{Constantinou-Young-2017,Constantinou-2018}. In the simpler quasi-geostrophic setup this transition to the upper branch is expected. One may obtain a lower bound for the volume-averaged zonal velocity and by imposing the extra restriction that the potential enstrophy power integral is balanced, one finds that the value of this lower bound increases with wind stress. Thus, for high enough wind stress values a transition to the upper branch must occur \cite{Constantinou-Young-2017}. The same transition occurs here in baroclinic runs. Regardless of whether the flow configuration is barotropic or baroclinic, the depth-averaged flow must obey this lower bound.

The depth-integrated momentum balance~\eqref{eq:totalmombal} helps us elucidate the bottom-layer westward flows in baroclinic configurations. With a flat bottom, topographic form stress vanishes and thus the bottom layer zonal flow has to be, on average, eastward. Thus, westward flows only appear with non-trivial bathymetry. Provided that topographic form stress is on average positive and balances most of the wind stress, the zonal flow can have either sign (since the sign of the zonal flow does {\color{black} not} affect the topographic form stress). The topographic form stress is determined by the time-mean bottom pressure (which is in turn determined by the the time-mean of all fluid interfaces above) and the mean bottom pressure can be configured so that topographic form stress is positive even if zonal flow is negative. (Similar westward bottom zonal flows were reported also, e.g., by \citeA{Treguier-McWilliams-1990} and \citeA{Stevens-Ivchenko-1997}. There is also some observational evidence of such bottom westward flows in certain regions of the Southern Ocean \cite{Cunningham-Barker-1996}.) Because of the strong sensitivity to bathymetry, we conclude that the development of bottom westward flow is not a robust feature of this stress-driven configuration.\footnote{A single ridge of the same height does not result in any bottom-layer westward flow (not shown here) {\color{black}and, therefore, the transport does not show any decrease for baroclinic configurations in regime~II.}).} These westward flows are, however,  responsible for the slight decrease in total transport with increasing wind stress that is observed in baroclinic configurations in flow regime~II of figure~\ref{fig:transport}.

\subsection{Time-mean--transient kinetic energy}\label{sec:resultdetails2}

The differences of the barotropic and the baroclinic runs are more pronounced when we inspect the time-mean--transient energy decomposition of the kinetic energy. Consider the standing--transient flow decomposition, $\bu = \overline{\bu}+\bu'$, where prime denotes fluctuations about the time-average. Figures~\ref{fig:transportdecomp}(e)-(h) depict the integrated mean kinetic energy~(MKE), $\sum_k \int\half \rhom |\overline{\bu}_k|^2\,\mathrm{d}x\,\mathrm{d}y$, and the eddy kinetic energy~(EKE),  $\sum_k \int \half\rhom \overline{|\bu'_k|^2}\,\mathrm{d}x\,\mathrm{d}y$, with $\rhom$ the mean reference density. What stands out is that for the barotropic cases, the MKE dominates over the EKE, while for baroclinic cases the opposite is true. For barotropic cases the eddies are negligible for wind stress values within flow regime~I (in which we have linear scaling for total transport) and within flow regime~IV (upper branch). {\color{black}Figures~\ref{fig:transportdecomp}(i)-(j) show the integrated mean available potential energy~(MAPE), $\sum_k \int \half \rhom g_{k-0.5}' \overline{\eta_{k-0.5}}^2\,\mathrm{d}x\,\mathrm{d}y$, and the eddy available potential energy (EAPE), $\sum_k \int \half \rhom g_{k-0.5}' \overline{\eta_{k-0.5}'^2}\,\mathrm{d}x\,\mathrm{d}y$, where $\eta_{k-0.5}$ is the deviation from rest of the fluid interface between layers $k$ and $k-1$ and $g'_{k-0.5}$ is the corresponding reduced gravity of that interface. The available potential energy is, in general, much larger than the corresponding kinetic energy. Furthermore, in both barotropic and baroclinic runs MAPE dominates over the EAPE.}

As we expect, the baroclinic cases have a  more vigorous eddy field, since baroclinic instability is very effective in diverting available potential energy to transient kinetic energy. Even for the weakly forced baroclinic cases in regime~I, after long enough spin up the top-layer fluid is accelerated to the point that the vertical shear is sufficiently large to render the flow baroclinically unstable. In contrast, figure~\ref{fig:transportdecomp}(e) shows that the barotropic runs with weak wind stress are steady without transients.  However, even for these weakly forced baroclinic cases, at equilibrium a time-mean flow must develop if topographic form stress is to balance most of the wind stress. It is interesting that the time-mean flow still accounts for about 25\% of the total kinetic energy (see figure~\ref{fig:snapshots}(h)). One might expect that the 25\% ratio is determined by the height of bathymetry. However, varying the height of the bathymetric features we found (not shown) that the ratio MKE$/$KE remains roughly constant until the bathymetry is small enough so that the topographic form stress fails to balance the wind stress and bottom drag takes over in~\eqref{eq:totalmombal}. For our channel, this transition occurs when bathymetric features are less than about 100$\,\text{m}$ tall (not shown).

\section{Discussion and Conclusions}

We outlined here a hierarchy of idealized models which have been used to investigate the relative contribution of baroclinic and barotropic processes in establishing zonal transport in a Southern Ocean-like channel. The models presented here are idealized in many respects (the geometry used, the simple bathymetry, constant wind stress, and the lack of diapycnal mixing or surface buoyancy forcing). Furthermore, the bathymetry we used does include any blocked geostrophic contours, $f/h$. {\color{black}Using bathymetry with blocked $f/h$ contours results in much less transient eddy activity for the 1-layer flow configurations. In addition, as \citeA{Masich-etal-2015} found, most topographic form stress in the Southern Ocean occurs when the flow meets continents, while our setup is restricted to relatively small ridges.} Regardless {\color{black}of these idealizations}, these models capture the basic processes involving the wind-driven component of the ACC.

A key outcome is that there exist parameter ranges (flow regimes~II~and~III) in which both barotropic and baroclinic configurations show similar time-mean zonal  transport values and also {\color{black}weak sensitivity} of the transport values to wind stress (figure~\ref{fig:transport}). This similarity in transport occurs despite dramatic differences in eddying flow (as can be seen, e.g., in figure~\ref{fig:snapshots} and figures~\ref{fig:transportdecomp}(e)-(h)). Our analysis shows that within flow regimes II~and~III the presence of transient eddies, regardless whether these eddies originate from baroclinic instability or other instabilities, renders the time-mean transport insensitive to wind stress forcing.

The similarity we find between the transport of the barotropic and the baroclinic configurations does not imply that the Southern Ocean is barotropic. Stratification and baroclinic mesoscale eddies are an integral element of the Southern Ocean; mesoscale eddies play a key role in overturning circulation, vertical heat transport and isopycnal tracer transport. However, the agreement in the values of the time-mean transport {\color{black}among flows with different layer configurations} despite differences in the eddy flow supports the notion that bathymetry actively shapes the standing eddy flow which in turn {\color{black}limits} the transport through the topographic form stress.

The notion that a regime resembling eddy saturation can be found in a barotropic fluid challenges the widely held view that eddy saturation is a consequence of baroclinic instability acting to optimize isopycnal slopes \cite{Straub-1993,Marshall-etal-2017}. The time-mean depth-integrated momentum balance (equation~\eqref{eq:totalmombal}) implies the mean topographic form stress is generated purely from the interaction between bathymetric slopes and the time-mean flow. This is easily seen by decomposing the bottom pressure into its standing and transient components, 
\vspace*{-0.2em}\begin{linenomath*}
\begin{equation}
\tfs = \big\langle\, \overline{\pb\,\partial_x{\hb}}\,\big\rangle = \big\langle\, \overline{\pb}\,\partial_x{\hb}\,\big\rangle.\vspace*{-0.2em}
\end{equation}
\end{linenomath*}
Thus, transient eddies can affect the net momentum balance only if they influence time-mean flow. For symmetric bathymetric features, like the Gaussian ridges we used here, a pressure field which is symmetric upstream and downstream a ridge does not result in mean topographic form stress. {\color{black}Friction alone is able to produce asymmetric flows over symmetric ridges, and therefore topographic form stress, even without transients. However, from the line of reasoning above}, we infer that transient eddies (generated either by baroclinic instability or barotropic processes) act to sharpen and barotropize flow over topography, leading to {\color{black}an even more} asymmetric flow over ridges \cite{Youngs-etal-2017}, which can enhance the net momentum sink due to topographic form stress.  Therefore, as an alternative to the common view that baroclinic instability is an integral component of eddy saturation, we propose that eddy saturation occurs as a consequence of feedbacks between transient eddies and the mean flow which creates topographic form stress and, in turn, balances the momentum input from wind stress (figure~\ref{fig:transportdecomp}\mbox{(c)-(d)}).

\vspace*{-0.5em}
\acknowledgments
We acknowledge fruitful discussions with Louis-Phillipe Nadeau, Callum Shakespeare, Kial Stewart, and William Young. {\color{black}We appreciate the constructive review comments by Dave Munday and an anonymous reviewer.} Furthermore, N.C.C.~is indebted to Alistair Adcroft, Angus Gibson, Bob Hallberg, and Aidan Heerdegen for their help regarding using MOM6. Numerical simulations were conducted on the Australian National Computational Infrastructure at ANU, which is supported by the Commonwealth of Australia. MOM6 code is available at \url{https://github.com/NOAA-GFDL/MOM6}. The model setup for the experiments presented can be found at \url{https://github.com/navidcy/EddySaturation-MOM6/}. Model output is available at \url{https://doi.org/10.5281/zenodo.3246030}.

\end{document}

More Information and Advice:

Math coded inside display math mode \[ ...\]
 will not be numbered, e.g.,:
 \[ x^2=y^2 + z^2\]

 Math coded inside \begin{equation} and \end{equation} will
 be automatically numbered, e.g.,:
 \begin{equation}
 x^2=y^2 + z^2
 \end{equation}

\begin{eqnarray}
  x_{1} & = & (x - x_{0}) \cos \Theta \nonumber \\
        && + (y - y_{0}) \sin \Theta  \nonumber \\
  y_{1} & = & -(x - x_{0}) \sin \Theta \nonumber \\
        && + (y - y_{0}) \cos \Theta.
\end{eqnarray}

